\documentclass[aps,onecolumn,notitlepage,superscriptaddress]{revtex4-2}

\usepackage[utf8]{inputenc}
\usepackage[english]{babel}
\usepackage[T1]{fontenc}
\usepackage[colorinlistoftodos]{todonotes}
\usepackage{graphicx}
\usepackage{amssymb,amsfonts,amsmath}
\usepackage{braket}
\usepackage{hyperref}
\usepackage{color}

\usepackage{soul}
\usepackage{comment}
\usepackage[normalem]{ulem}
\usepackage{enumitem}

\newcommand{\I}[0]{\mathcal{I}}
\newcommand{\M}[0]{\mathcal{M}}

\newcommand{\inp}[0]{\mathrm{in}}
\newcommand{\out}[0]{\mathrm{out}}

\begin{document}

\title{Unraveling quantum phase estimation: exploring the impact of multi-photon interference on the quantum Fisher information}

\author{A. Ma}
\affiliation{Departamento de F\'{\i}sica, FCEyN, UBA. Ciudad Universitaria, 1428 Buenos Aires, Argentina}
\author{A. G. Magnoni}
\affiliation{Departamento de F\'{\i}sica, FCEyN, UBA. Ciudad Universitaria, 1428 Buenos Aires, Argentina}
\affiliation{CITEDEF \& UNIDEF - CONICET, J.B. de La Salle 4397, 1603 Villa Martelli, Buenos Aires, Argentina}
\author{M. A. Larotonda}
\affiliation{Departamento de F\'{\i}sica, FCEyN, UBA. Ciudad Universitaria, 1428 Buenos Aires, Argentina}
\affiliation{CITEDEF \& UNIDEF - CONICET, J.B. de La Salle 4397, 1603 Villa Martelli, Buenos Aires, Argentina}
\author{L. T. Knoll}
\affiliation{Departamento de F\'{\i}sica, FCEyN, UBA. Ciudad Universitaria, 1428 Buenos Aires, Argentina}
\affiliation{CITEDEF \& UNIDEF - CONICET, J.B. de La Salle 4397, 1603 Villa Martelli, Buenos Aires, Argentina}

\email{Corresponding author: lknoll@citedef.gob.ar}

\begin{abstract}

Quantum interference is known to become extinct with distinguishing information, as illustrated by the ubiquitous double-slit experiment or the two-photon Hong-Ou-Mandel effect. 
In the former case single particle interference is destroyed with which-path information while in the latter bunching interference tails-off as photons become distinguishable.
It has been observed that when more than two particles are involved, these interference patterns are in general a non-monotonic function of the distinguishability.
Here we perform a comprehensive characterization, both theoretically and experimentally, of four-photon interference by analyzing the corresponding correlation functions, contemplating several degrees of distinguishability across different parameters. 
This study provides all the necessary tools to quantify the impact of multi-photon interference on precision measurements of parameters such as phase, frequency, and time difference.
We apply these insights to quantify the precision in the estimation of an interferometric phase in a two-port interferometer using a four-photon state. 
Our results reveal that, for certain phase values, partially distinguishable multi-photon states can achieve higher Fisher information values compared to the two-photon experiment.
These findings highlight the potential of distinguishable multi-photon states for enhanced precision in quantum metrology and related applications.

\end{abstract}

\date{\today}
\maketitle

\section{Introduction}

Photon-based quantum technologies rely on
quantum interference as an essential resource
for applications such as quantum computation, simulation, and quantum metrology \cite{Zhong2020,Polino2020,obrien2009}.
Their implementation is heavily based on the development of sources of identical single photons, where indistinguishability between photons is a desirable feature \cite{Ding2016,Tomm2021}. 
{These sources enable precise control over quantum states, which is essential for the aforementioned applications as well as secure quantum communication and high-resolution quantum imaging \cite{couteau2023applications, defienne2024advances, moreau2019imaging}.
Indistinguishability between photons enables the creation of entangled states and allows the execution of complex quantum algorithms with high fidelity \cite{hu2023progress, pirandola2018advances}.
In quantum metrology, indistinguishable photons enhance measurement precision and sensitivity. This capability is fundamental for developing ultra-sensitive sensors and improving standards of measurement \cite{mitchell2004super, demkowicz2015quantum,Yoon2024}.
}
As quantum systems increase in scale, { interference between indistinguishable photons may arise, and understanding its effect in quantum metrological tasks} is crucial from a fundamental as well as a technological point of view.

Indistinguishability and quantum interference have gone hand in hand when dealing with single particles and two-particle systems, given that information allowing distinguishability of a system can destroy interference.
In the double-slit experiment, single particles may lead to an interference pattern as long as the which-path information is not available \cite{Feynman,scully1991,mandel1991}. When an observer can tell with certainty which path the particle took, interference is completely destroyed. 
The Hong-Ou-Mandel (HOM) interference effect shows that when two identical particles impinge on the input faces of a 50:50 beam splitter, both photons exit the beam splitter together through the same output \cite{hong1987}. This bunching effect vanishes monotonically as the photons become distinguishable in any degree of freedom. {In fact, HOM interferometry is the standard setup for the characterization of indistinguishability.} 
Distinguishing information can arise from interactions with the environment or decoherence effects that ultimately destroy quantum interference, thus undermining any quantum advantage. 

For more than two photons, interference is in general a non-monotonic function of the distinguishability \cite{Tichy2011,Ra2013a} and different experiments have shown that quantum interference does not always vanish with distinguishability \cite{jones2020,messen2017,jones2023}, 
proving that multi-particle interference is substantially different than the two-particle case
as multi-particle interference may persist even in the presence of distinguishing information. 
Understanding multi-photon interference is therefore of great relevance for the development of quantum technologies relying on photon interference.
{In particular, for quantum metrological tasks such as phase estimation, characterizing the effect of multi-photon interference and distinguishability may allow a better understanding on the achievable precision.
In the problem of phase estimation, more photons usually imply better precision. 
Without quantum correlations, the minimum phase uncertainty $\Delta\phi$ achievable using $N$ photons is given by the \emph{standard quantum limit} (SQL), where $(\Delta\phi)\sim\frac{1 }{\sqrt{N}}$.
Quantum metrology, however, allows us to improve the precision: by using phenomena such as entanglement, it is possible to reach the \emph{Heisenberg limit} (HL). This limit defines the ultimate bound on the uncertainty: $(\Delta\phi)\sim\frac{1}{N}$ \cite{giovannetti2004quantum,Toth2014}.
In this way, using multi-photon states in a particularly engineered way, phase measurements with an uncertainty below the SQL can be obtained \cite{ligo2011gravitational,dowling1998correlated,qin2023unconditional,ou2020quantum,Krischek2011}. Achieving this ultimate bound 
requires the preparation of highly entangled states such as the $N00N$ state \cite{Bollinger1996,Walter2004,Nagata2007,Afek2010}, a process that strongly relies on indistinguishability, making it essential to harness the full potential of quantum technologies and achieve breakthroughs in quantum metrology \cite{Birchall2016,Jachura2016,knoll2019}.
However, creating $N00N$ states with a large number of photons $N$ is challenging, and the maximum attainable precision becomes highly susceptible to imperfections and photon losses \cite{Berchera2019,Roccia2018,Birchall2020,Vidrighin2014,Namkung2024}.

In this work, we perform a characterization of four-photon interference in a two-port interferometer. We analyze how distinguishability affects phase estimation by evaluating its impact on the attainable precision, quantified by the Fisher information.
We study the correlation functions associated with our particular scheme, contemplating several degrees of distinguishability across different parameters. 
Our theoretical model may be used to evaluate the precision of parameters such as phase, frequency, and time difference. We focus particularly on the problem of phase estimation and perform a photon-based experiment to test our model. Instead of using a four-photon $N00N$ state, a Holland-Burnett state is used as input state, which is less challenging experimentally but still achieves high sensitivity \cite{HB}. 

The paper is organized as follows. In Sec. \ref{sec:theory} we briefly present the theory of phase estimation, introducing the definitions of basic quantities such as Fisher information, and the key concepts for calculating the correlation functions that describe the probability distributions for photon interference. We then describe our particular scheme for phase estimation (Sec. \ref{sec:theo_scheme}) and calculate the probability distributions. The achievable precision is then obtained by means of the Fisher information in Sec. \ref{sec:theo_qfi}. In Sec. \ref{sec:exp&res} we describe the experimental setup and the results obtained. Finally, in Sec. \ref{sec:conc} we present the conclusions. Appendix \ref{app:A} contains a detailed description of the calculation of the probability distributions derived in Sec. \ref{sec:theo_scheme}.}

\section{Interference and phase estimation}

\subsection{Theory}
\label{sec:theory}

{Phase estimation is an archetypic problem in quantum metrology. The estimation of an optical phase shift represents many physical problems, as it can be related to different experimental quantities such as distance, velocity, material properties \cite{prop1,prop2}, or even the presence of gravitational waves \cite{aasi2013enhanced}. The problem consists on estimating an unknown phase shift between two different modes inside an interferometer. These modes can be represented by different degrees of freedom of the photons, such as path, polarization, or orbital angular momentum. 
In a general scheme for quantum phase estimation, the parameter is encoded in a probe state $\rho^\inp$ by means of a unitary operation $U_\phi$ applied to the initial state such that $\rho^\out(\phi)=U_\phi\rho^\inp U_\phi^{\dagger}$ (we will restrict our analysis to unitary evolutions but it is worth to note that this procedure can be extended to the more general case of non-unitary maps). Information about the parameter is retrieved after performing a measurement on $\rho^\out(\phi)$ that is, in general, given by a positive-operator valued measure (POVM) $\M=\{M_m\}$ with possible outcomes $m$. Finally, an estimator $\hat{\phi}$ is obtained from the outcome probabilities $p(m|\phi)$.

}
{
 
The lower bound for the variance of the estimated parameter $\hat{\phi}$ is given by the quantum Cram\'er-Rao bound~\cite{Helstrom1967,Holevo1982} (QCRB):
\begin{equation}
\label{eq:QCRB}
\Delta^2(\hat{\phi}) \geq \frac{1}{F_Q(\phi)},
\end{equation}
where $F_Q(\phi) = \max_{\{\M\}} F(\phi)$ is the quantum Fisher information (QFI), and $F(\phi)$ the Fisher information (FI) defined as:
\begin{equation}
\label{eq:FI}
F(\phi)=\sum_m \frac{1}{p(m|\phi)}\left|\frac{dp(m|\phi)}{d\phi}\right|^2.
\end{equation}
The QFI is then determined from the corresponding probabilities $p(m|\phi)$
and the maximum is taken over all possible POVMs ${\{\M\}}$.
Therefore, the QFI quantifies the maximal information
on the parameter to be estimated.
For the particular case of phase estimation from the unitary evolution $U_\phi = \exp(-i\phi H)$, 
and for {an arbitrary} initial pure state $\rho^\inp = \ket{\Phi^\inp}\bra{\Phi^\inp}$, the QFI has a closed expression and does not depend on $\phi$. It is obtained from the variance of the Hamiltonian in the initial state \cite{Paris2009,Toth2014}:
\begin{equation}
\label{eq:QFI}
F_Q(\ket{\Phi^\inp})= 4 \Delta^2(H) =4 \left [\braket{\Phi^\inp |H^2|\Phi^\inp} - \left(\braket{\Phi^\inp |H|\Phi^\inp}\right)^2\right ]. 
\end{equation}

In this work, we calculate the precision of a phase estimation problem in a two-mode interferometer using a four-photon input state.
To this end, we will use the FI to evaluate the sensitivity of the proposed scheme and benchmark it with the QFI. The probabilities $p(m|\phi)$ needed to evaluate the FI can be calculated by describing the interference between photons. }

In the simpler, two-photon case, interference is determined by two photons labeled with arrival times $t_1$ and $t_2$ at the detector, and the detection probability can be found using the average second order correlation function, defined as:
\begin{equation}
    G^{(2)}(t_1,t_2) =tr[\hat g\rho^\inp],
\end{equation}
where $\rho^\inp$ is the initial state of the two-particle system, and $\hat g$ is the operator that describes the two-particle measurement.
In this way, we can define the joint probability of a photodetection in two detectors as:
\begin{equation}
    P^{(2)}(\delta\tau_1,\delta\tau_2,\Delta t_0)=\int_{\delta\tau_1-\Delta t_0/2}^{\delta\tau_1+\Delta t_0/2}dt_1\,\int_{\delta\tau_2-\Delta t_0/2}^{\delta\tau_2+\Delta t_0/2}dt_2\,G^{(2)}(t_1,t_2),
\end{equation}
where the integration limits correspond to the detection interval for each particle with a relative time difference $|\delta\tau_1-\delta\tau_2|$ between both intervals, and a detection window at each detector given by $\Delta t_0$. 

In realistic experimental scenarios, the resolution window is not sufficiently narrow to distinguish each wave packet precisely, such that $\Delta t_0\gg t_c$, with $t_c$ the coherence time of the wave packet. Then, the integration limits can be extended to infinity \cite{legero}:
\begin{equation}
   P^{(2)}=\iint_{-\infty}^{\infty} dt_1dt_2\,G^{(2)}(t_1,t_2)\equiv\iint dt_1dt_2\,G^{(2)}(t_1,t_2).
    \label{ec.probabilidad dos fotones-HOM camino}
\end{equation}    

The same argument can be applied to the case of four photons, where in the approximation of $\Delta t_0\gg t_c$ the four-photon detection probability can be expressed as:

\begin{equation}
    P^{(4)}=\iiiint dt_1dt_2dt_3dt_4\,G^{(4)}(t_1,t_2,t_3,t_4).
    \label{ec.probabilidad cuatro fotones-HOM camino}
\end{equation}

To calculate this probability, we need to obtain the explicit expression for the fourth-order correlation function $G^{(4)}$, which is specific to the designed scheme for phase estimation {described in the following subsection}.

\subsection{{Scheme for} four-photon interference}
\label{sec:theo_scheme}
Our model consists on an interferometric system composed of three parts: a photon pair source, a polarization-based Hong-Ou-Mandel interferometer and a detection system. 
We will focus on the case of $N$ photons entering a HOM interferometer such that the input state is described by: 
\begin{equation}
    \ket{\Phi^\inp}=\ket{N/2,N/2}.
\end{equation}
That is, $N/2$ photons impinge on each of the two input ports of a balanced beam splitter. We will show below that this configuration allows for the generation of a particular type of output state, which is the Holland-Burnett state \cite{HB}
\begin{equation}
    \ket{\Psi}_{HB}= \sum_{n=0}^{N/2}c_n\ket{2n,N-2n},\quad c_n=\frac{\sqrt{(2n)!(N-2n)!}}{2^{N/2}n!(N/2-n)!}.
    \label{ec.HB} 
\end{equation}
This $N$-photon state achieves a high phase sensitivity, as it will be detailed in the next section. We will restrict our analysis and experiment to the case $N=4$. Note that for the case of $N=2$ the state described in Eq. \eqref{ec.HB} results in a two-photon $N00N$ state.

\begin{figure}[h!]
    \centering
       \includegraphics[width=0.4\textwidth]{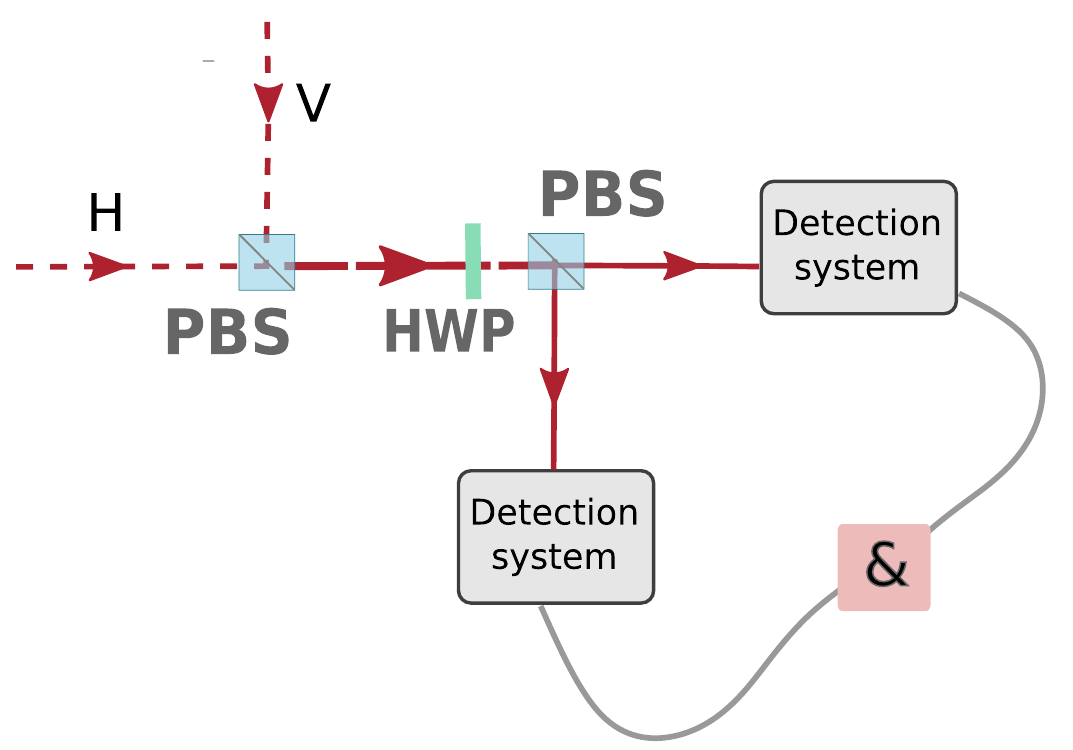}
    \caption{General scheme describing the setup for characterizing the output of the polarization HOM interferometer: Distinguishable polarization modes are combined in a common path by means of a first PBS. A half wave plate erases the information in the polarization degree of freedom and a second PBS projects the two polarization modes for detection.}
    \label{fig:HOMpol}
\end{figure}

As mentioned above, we will consider a polarization-based HOM interferometer consisting of a polarizing beam splitter (PBS) and a half-wave plate (HWP) followed by another PBS, as shown schematically in Fig. \ref{fig:HOMpol}. 

The polarization HOM interferometer is an alternative version in which polarization modes $H$ (horizontal) and $V$ (vertical) play the same role as path modes in the traditional HOM interferometer. Photons with orthogonal polarization travel in orthogonal directions towards the two input faces of the PBS. Due to the nature of the PBS, the horizontally polarized photons will be transmitted, and the vertically polarized ones will be reflected. This means that the two beams are combined into the same output path. Immediately after the PBS, a half-wave plate is placed on the beam path, allowing the rotation of the polarization state of the incoming photons, and adding an arbitrary phase difference between the two polarization modes. Finally, a second PBS is placed after the HWP. Photons are detected at each output of the PBS with single photon detectors.

Therefore, the first PBS allows for the combination of both beams in a single path. In this interferometer, the path modes are assumed indistinguishable (both beams are combined into a single path by the first PBS), and interference can be obtained in the polarization degree of freedom with the combination of the HWP followed by the second PBS.
Under the assumption that the PBS is balanced, the transformation of the creation operators $\hat a^\dag$ in each mode is uniquely defined by the angle of the HWP:
\begin{equation}
    \begin{pmatrix}
    \hat a^\dag_{s_3}\\
    \hat a^\dag_{s_4}
\end{pmatrix}=-i\begin{pmatrix}
    \cos(2\theta)&\sin(2\theta)\\
    \sin(2\theta)&-\cos(2\theta)
\end{pmatrix}\begin{pmatrix}
    \hat a^\dag_{s_1}\\\hat a^\dag_{s_2}
\end{pmatrix},
\label{ec:HWP}
\end{equation}
where modes $s_1,s_2,s_3,s_4$ describe linear polarization states. The phase to be estimated, $\phi$, is encoded in the physical angle of the HWP $\theta$, such that $\theta=\phi/4$.
For the particular case of $\theta=\pi/8$ ($\phi=\pi/2$), the matrix in Eq. \eqref{ec:HWP} is equivalent to that of a balanced beam splitter.
Then, by adjusting the angle of the half-wave plate, it is possible to introduce a relative phase difference between the polarization modes of the photons in the input state, generating an interference pattern.

We will assume that photons are generated in pairs. In the simplest case, the four photons arrive at the HWP at the same time, two photons with polarization $s_1$ and other two photons with polarization $s_2$.
In a situation more closely resembling the experimental conditions, the photons can differ in their spatiotemporal and spectral modes; then the photon wave packet operator formalism is used to find the probability distribution of each state {(see Appendix \ref{app:A})}.
After interacting with the first PBS, the four photons are combined in a single path mode and the initial state 
\begin{equation}
\ket{\Phi^\inp}=\ket{2_{\xi_1}}_{s_1}\ket{2_{\xi_2}}_{s_2}
    \label{ec.estado inicial pol-4 fotones}
\end{equation}
is generated, where two photons are in mode $\xi_1$ with polarization $s_1$ and the other two photons are in mode $\xi_2$ with polarization $s_2$.

The probability distributions are then described by the aforementioned correlation functions, with the corresponding observables.
The general expression for the fourth order correlation function is given by:

\begin{equation}
    G^{(4)}(t_1,t_2,t_3,t_4)=\sum_{s,s',s'',s'''}tr\left[\hat A_{s,s',s'',s'''}(t_1,t_2,t_3,t_4)\rho^\inp\right],
\end{equation}
where $\hat A(t_1,t_2,t_3,t_4)$ is the temporal correlation operator that describes the arrival of photons with polarization along arbitrary directions $s,s',s'',s'''$ to the detectors at each time $t_1,t_2,t_3,t_4$ respectively, and can be expressed in terms of the photon wave packet operators:
\begin{widetext}
\begin{equation}
    \hat A_{s,s',s'',s'''}(t_1,t_2,t_3,t_4)=\hat a_{s}^\dag(t_1)\hat a_{s'}^\dag(t_2)\hat a_{s''}^\dag(t_3)\hat a_{s'''}^\dag(t_4)\hat a_{s'''}(t_4)\hat a_{s''}(t_3)\hat a_{s'}(t_2)\hat a_{s}(t_1).   
\end{equation}
\end{widetext}

This operator is different for each detection configuration at the output of the second PBS, where photons can be distributed in each of the two output faces of the PBS according to their polarization.
We can differentiate 3 cases, with the following operators:
\begin{itemize}
    \item $\hat A^{4:0}_{s_3,s_3,s_3,s_3}$ where all four photons are polarized in the direction $s_3$. This term will contribute to the probability of detecting state $\ket{4}_{s_3}\ket{0}_{s4}\equiv\ket{4,0}$. The case of state $\ket{0,4}$ is equivalent.

    \item $\hat A^{3:1}_{s_3,s_3,s_3,s_4}$ where three photons exit the PBS through the same path mode with their polarization oriented along direction $s_3$, while one photon exits the PBS through the other path mode with polarization direction $s_4$, orthogonal to $s_3$. This operator describes the probability distribution associated to state $\ket{3,1}$ (case $\ket{1,3}$ is equivalent).
    
    \item $\hat A^{2:2}_{s_3,s_3,s_4,s_4}$ where two photons are polarized along direction $s_3$ at the output of the PBS, while the other two photons are oriented along the orthogonal polarization direction $s_4$ at the other output path of the PBS. This observable describes the probability of detecting the state $\ket{2,2}$.
    
\end{itemize}

\begin{figure*}
    \centering
       \includegraphics[width=0.95\textwidth]{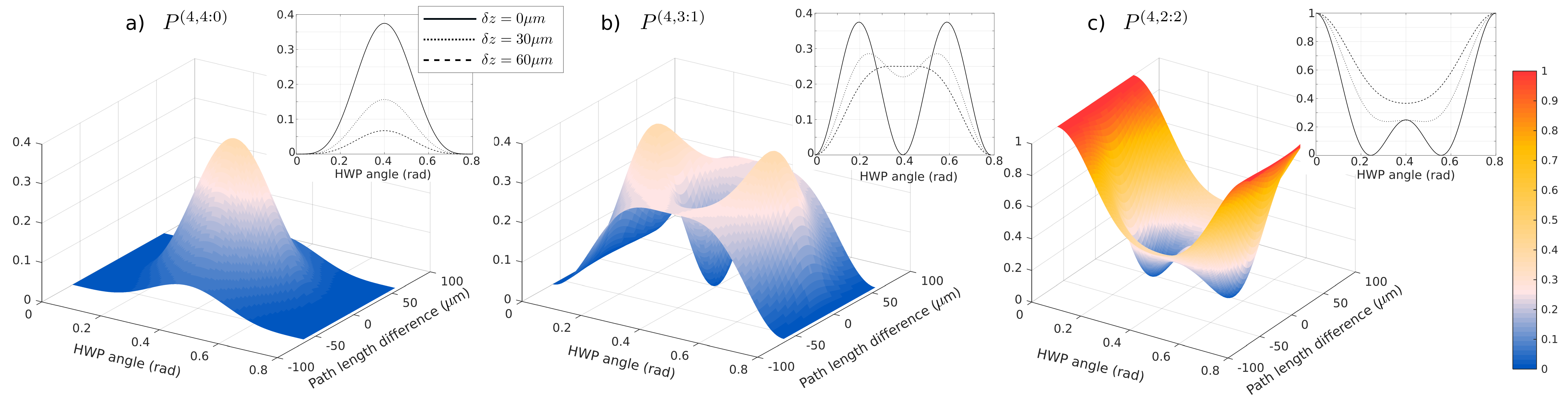}
    \caption{Probability distributions a) $P^{(4,4:0)}$; b) $P^{(4,3:1)}$ and c) $P^{(4,2:2)}$ as a function of the phase difference given by the angle of the HWP $\theta$ on the $x$ axis and the path length difference $\delta z$ on the $y$ axis. The insets correspond to {cuts in the $y$ axis} for three particular cases of $\delta z=0 \mu m$ (solid line); $\delta z=30 \mu m$ (dotted line) and $\delta z=60 \mu m$ (dashed line).}
    \label{fig:proba}
\end{figure*}

All these operators can be expressed in terms of the input operators using Eq. \eqref{ec:HWP}. In this way, each probability distribution can be calculated, obtaining (see Appendix \ref{app:A} for a detailed calculation):
%
\begin{subequations}
\begin{align}
    &\begin{aligned}\begin{split}
        P^{(4,4:0)}&(\theta, \delta\tau,\delta \omega)=\cos^4(2\theta)\sin^4(2\theta)\left(1+ e^{-\frac{8\delta\tau^2}{ t_c^2}-2\,\delta \omega^2 t_c^2}+4e^{-\frac{4\delta\tau^2}{ t_c^2}-\delta \omega^2 t_c^2}\right),
    \end{split}\end{aligned}\label{ec.probabilidad(4,4:0)-HOM pol}\\
    &\begin{aligned}\begin{split}
        P^{(4,3:1)}&(\theta, \delta\tau,\delta \omega)=2\cos^2(2\theta)\sin^2(2\theta)\Bigg(\sin^4(2\theta)+\cos^4(2\theta)+\\
        &+2(\sin^2(2\theta)-\cos^2(2\theta))^2e^{-\frac{4\delta\tau^2}{ t_c^2}-\delta \omega^2 t_c^2}-2\cos^2(2\theta)\sin^2(2\theta)e^{-\frac{8\delta\tau^2}{ t_c^2}-2\,\delta \omega^2 t_c^2}\Bigg),
    \end{split}\end{aligned}\label{ec.probabilidad(4,3:1)-HOM pol}\\
    &\begin{aligned}\begin{split}
        P^{(4,2:2)}&(\theta, \delta\tau,\delta \omega)=\sin^8(2\theta)+4\cos^4(2\theta)\sin^4(2\theta)+\cos^8(2\theta)+\\
        &+8\left(\sin^4(2\theta)\cos^4(2\theta)-\cos^6(2\theta)\sin^2(2\theta)-\sin^6(2\theta)\cos^2(2\theta)\right)e^{-\frac{4\delta\tau^2}{ t_c^2}-\delta \omega^2 t_c^2}+\\
        &+6\sin^4(2\theta)\cos^4(2\theta)e^{-\frac{8\delta\tau^2}{ t_c^2}-2\,\delta \omega^2 t_c^2},
    \end{split}\end{aligned}\label{ec.probabilidad(4,2:2)-HOM pol}
\end{align}
\label{ec:probabilidad(4)-HOM pol}
\end{subequations}
%
with $\delta\omega=(\omega_2-\omega_1)/2$ and $\delta\tau = (\tau_2-\tau_1)/2$, being $\tau_{1,2}$ the times at which each photon pair arrives at the HWP.
To better understand these distributions, Fig. \ref{fig:proba} shows each one as a function of the phase difference given by the angle of the HWP $\theta$ and the path length difference $\delta z$, where we used $\delta\tau=\delta z/c$. In all cases we assumed $\delta\omega=0$, that is, frequency indistinguishability, and $t_c=l_c/c$ with $l_c=60\mu m$ the coherence length of the photon wave packet. {These conditions were chosen based on the accessible experimental parameters and describe a situation where the angle of the HWP represents the phase to be estimated, while the path length difference controls the distinguishability between photons. The assumption $\delta\omega=0$ is justified by the use of narrowband interference filters. More details on the experimental scheme will be given in Sec. \ref{sec:exp&res}.}

From Fig. \ref{fig:proba}, it can be observed that the probability distributions exhibit a high sensitivity to variations in the parameter $\theta$. 
In particular, for indistinguishable photon paths ($\delta z=0$), $P^{(4,4:0)}$ achieves its maximum value for $\theta=\pi/8$, while $P^{(4,3:1)}$ reaches a minimum zero value, contributing to the four-photon bunching interference. 
In analogy to the two photon HOM effect, we would also expect $P^{(4,2:2)}$ to have a minimum at this point.
On the contrary, $P^{(4,2:2)}$ shows an enhancement at $\theta=\pi/8$ and two (minimum) zero values located symmetrically with respect to this point. 
This effect has already been reported \cite{Tichy2011,Ra2013a}, proving that when four particles interfere simultaneously there is a non-monotonic transition as the photons become distinguishable. In our case, this behavior arises as the polarization modes become orthogonal, even though the photon paths are indistinguishable. The non-monotonic behavior can also be seen for $P^{(4,3:1)}$ as a function of the phase difference.
{An interesting effect that can be observed is that the interference patterns behave differently if the distinguishing information arises in the polarization or in the path degree of freedom. For completely distinguishable photon paths, we can still observe a phase modulation, while for orthogonal polarization modes interference vanishes completely in the path degree of freedom. Also, for complete indistinguishability in the photon's path, the non-monotonic behavior is present in both $P^{(4,3:1)}$ and $P^{(4,2:2)}$, while for indistinguishable polarization modes this non-monotonic transition is only seen for $P^{(4,2:2)}$.} As the path length difference increases, the pattern close to $\theta = \pi/8$ experiences fewer fluctuations and becomes more uniform, implying that the system loses the ability to record changes in $\theta$.
We will use the Fisher information to describe how the probability distributions in Eq. \eqref{ec:probabilidad(4)-HOM pol} affect the precision achievable in the problem of phase estimation, and to quantify the sensitivity of interference mentioned above.

\subsection{Fisher information}
\label{sec:theo_qfi}

\begin{figure}[h!]
    \centering
       \includegraphics[width=0.45\textwidth]{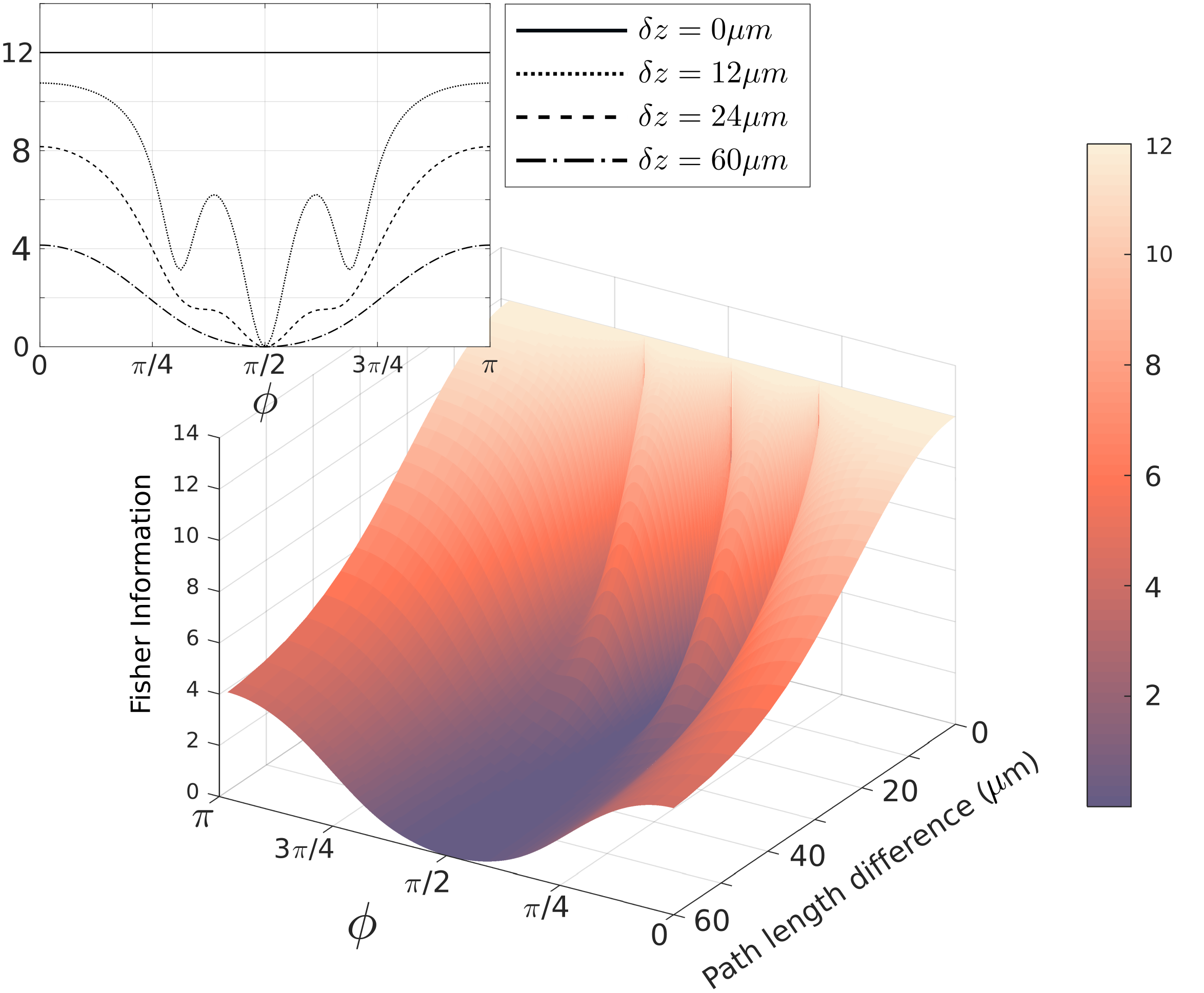}
    \caption{Fisher information as a function of the phase difference $\phi$ and the path length difference $\delta z$. For indistinguishable photons ($\delta z$=0) a maximum value of $F=12$ is achieved for all phase values. With increasing indistinguishability (as the path length difference increases) the FI decreases and becomes dependent on the phase. This can be seen on the inset for four different values of $\delta z$.}
    \label{fig:FI_4}
\end{figure}

As it was mentioned above, the goal is to use the FI to evaluate the achievable precision of phase estimation inside a two-port interferometer, using a Holland-Burnett state with a definite number of photons $N$ as the input state. It has been shown \cite{Eaton2021} that for the input state given by Eq. \eqref{ec.HB}, that is, $\ket{\Phi^\inp}=\ket{\Psi_{HB}}$ the QFI is given by 
\begin{equation}
    F_Q(\ket{\Psi_{HB}},N)=N(N/2+1),    
    \label{eq:QFI_HB}
\end{equation} 
such that the variance scales as the Heisenberg limit ($\Delta^2(\hat\phi)\sim 1/N^2$).
{This ultimate bound is obtained using Eq. \eqref{eq:QFI}. Achieving this precision in an experimental scenario implies performing a series of measurements that maximize the FI. In our case, where the POVM was chosen from the experimentally accessible measurement conditions, the precision can be calculated from the FI and compared to the QFI. Therefore,}
to evaluate the phase sensitivity in our scheme, we use the Fisher information defined as:
\begin{equation}
    F(\phi)=\sum_m\frac{1}{P^{(4,m)}(\phi)}\left(\frac{\partial P^{(4,m)}(\phi)}{\partial \phi}\right)^2
    \label{eq:FI_classic}
\end{equation}
with $m\in\left\{4:0,0:4,3:1,1:3,2:2  \right\}$ and the probabilities given by Eq. \eqref{ec:probabilidad(4)-HOM pol}, with $\theta=\phi/4$.
The Fisher information in Eq. \eqref{eq:FI_classic} can be calculated as a function of the phase difference $\phi\in [0,\pi]$ for different values of the path length difference $\delta z\in [0,l_c]$ (see Fig. \ref{fig:FI_4}). 
For complete indistinguishability ($\delta z=0$) we obtain $F(\phi)\equiv 12 \quad\forall\quad\phi$, which is exactly the value of the QFI in Eq. \eqref{eq:QFI_HB} for $N=4$: $F_Q(\phi)=12$. Therefore, for indistinguishable photons, the maximum precision is achieved for the selected POVM and it is independent of the parameter to be estimated.
 
Increasing the path length difference results in a decrease in the Fisher information; also, a dependence with the phase appears. 
When $\delta z \geq l_c$ the Fisher information reaches the SQL ($F(\phi)=4$) for certain phase values, as it can be seen in the inset of Fig. \ref{fig:FI_4}. 
For all values of $\delta z \neq 0$ the Fisher information goes to zero for $\phi=\pi/2$.
As mentioned before, when the path difference increases and the phase difference is close to $\phi=\pi/2$ ($\theta = \pi/8$) the system becomes less sensitive to changes in the phase, losing precision in its estimation.

An interesting feature that can be seen from these figures is that for indistinguishable photon paths ($\delta z=0$), the probability distributions present a high sensitivity to variations in the phase, while the FI remains constant and equal to the ultimate value given by the QFI for our particular scheme, even in the presence of a non-monotonic behavior in the phase as the one given by $P^{(4,2:2)}$ or $P^{(4,3:1)}$.
This suggests that the enhancement on the precision is achieved by interference patterns exhibiting significant variations in the phase.

For the particular case of $N/2$ photons per input mode in a two-port interferometer it has been shown that the QFI increases linearly with respect to the degree of
indistinguishability  $0\le \mathcal{I} \le 1$ (controlled by the path length difference in our setup), and does not depend on the phase $\phi$ to be estimated: $F_Q(\phi)= N\left(\frac{N}{2} \I+1\right)$ \cite{knoll2019,knoll2023}. Hence, for a given POVM, we can obtain a FI that is independent of the phase for any degree of indistinguishability. 
The chosen POVM in our scheme does not attain the QFI, except for $\delta z=0$ ($\I=1$, i.e., complete indistinguishability). Nonetheless, for certain phase values, the multi-photon state enables higher values of FI compared to the QFI in the best two-particle scenario, achieving higher precision in the estimation process.

\section{Experiment and Results}
\label{sec:exp&res}

\begin{figure}[h!]
    \centering
       \includegraphics[width=0.75\textwidth]{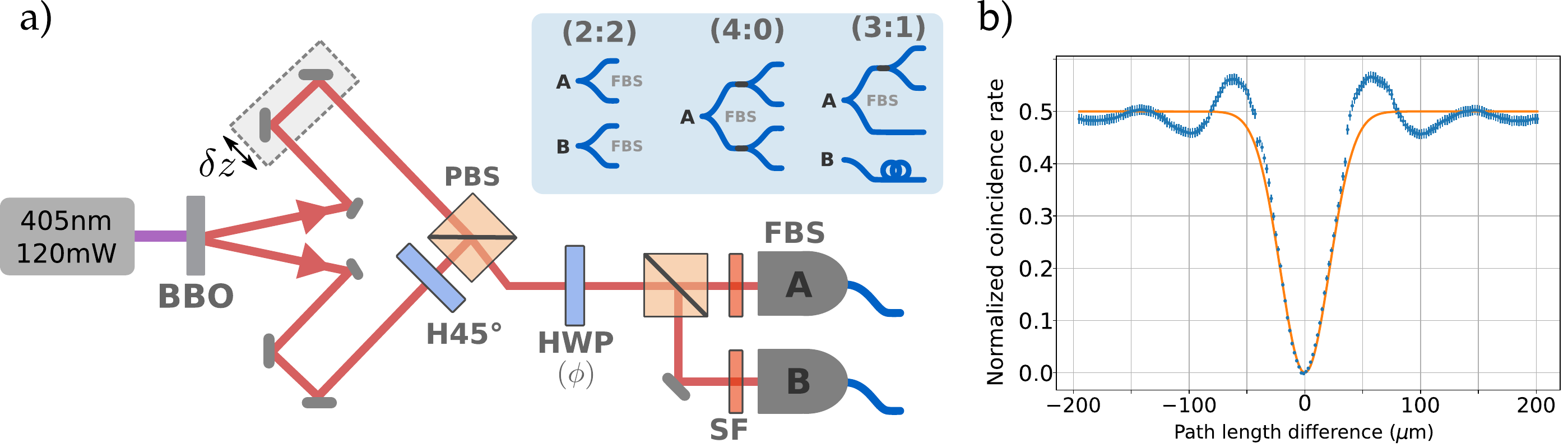}
    \caption{a) Experimental setup consisting of a photon pair source generating horizontally polarized photons by SPDC. These photons are then recombiend into a single path mode by means of a HWP at 45$^{\circ}$ and a PBS. Polarization interference is then achieved by a HWP adding a phase difference $\phi$ between both polarization modes. Photons are then projected by a PBS into the output modes and coupled to the FBSs A and B, which are assembled in different configurations associated to each probability measurement (outlined in the blue inset). The photons bandwidth is defined by the use of spectral filters (SF). b) HOM interference: Two-fold coincidences as a function of the path length difference. The blue dots correspond to the normalized coincidence counts after background subtraction. Error bars were calculated assuming the count rates follow a Poisson distribution. The solid line corresponds to the fitted two-photon coincidence probability distribution. {For a path length difference $\delta z=0$ we obtain a photon indistinguishability of $\mathcal{I}=0.998\pm0.001$}.}
    \label{fig:setup}
\end{figure}

As described in Sec. \ref{sec:theo_scheme}, our experimental scheme (Fig. \ref{fig:setup}a)) consists of three parts: i) a photon pair source;  ii) a polarization based interferometer; iii) a detection system. Photon pairs are generated by Spontaneous Parametric Down Conversion (SPDC) in a BBO type-I nonlinear crystal pumped by a CW 405 nm vertically polarized laser (120 mW). Pairs of horizontally polarized photons are generated at 810 nm and directed into two distinct path modes.

Our photon pair source has an effective coincidence/single count ratio of $0.30\pm 0.02$, with a maximum two-fold coincidence count rate of 150 kHz. The measured four-fold coincidence rate was 45 Hz. These coincidences were obtained using fiber beam splitters (FBS) coupled to each path, four single photon detectors (Si APD) and an FPGA-based coincidence counter. Interference filters with $\Delta\lambda =10$nm (FWHM) were placed before the coupling to each FBS. The coincidence time window was 20 ns, determined by the FPGA's internal clock.

The generated photons in each of the two paths are then recombined into a single path mode using a HWP at $45^{\circ}$ in one of the paths, that rotates the horizontally polarized photons into vertically polarized ones, followed by a PBS. In this way, the vertically polarized photons are reflected by the PBS and the horizontally polarized photons on the other path are transmitted into the same output path after the PBS (see Fig. \ref{fig:setup}a)). 
Subsequently, a polarization-based interferometer is implemented as described in Sec. \ref{sec:theo_scheme}. 
Each probability distribution in Eq. \eqref{ec:probabilidad(4)-HOM pol} is detected individually using FBSs in a tree configuration in order to resolve each contribution (see inset in Fig. \ref{fig:setup}a)).

By registering the two-fold coincidences as a function of the path length difference for $\phi = \pi/2$ ($\theta = \pi / 8$), the traditional two-photon HOM interference effect can be observed and used as a measure of indistinguishability. Fig. \ref{fig:setup}b) 
shows the normalized coincidence counts as a function of the path
length difference (blue dots). The solid line corresponds to the
fitted two-photon coincidence probability distribution \cite{legero}. The obtained visibility, which defines the degree of indistinguishability, is $\mathcal{I}= 0.998 \pm 0.001$.
{The location of the dip determines the condition for complete path indistinguishability ($\delta z=0$). The width of the dip is a measure of the coherence length of the photon wave packet, which in turn is determined by the imposed spectral bandwidth. For any path length difference greater than $l_c$, complete distinguishability is achieved.} The observed oscillations outside the HOM dip correspond to the spectral filtering given by the use of rectangular-shaped bandwidth transmission interference filters, in addition to the fiber coupling.

\begin{figure}[h!]
    \centering
       \includegraphics[width=0.8\textwidth]{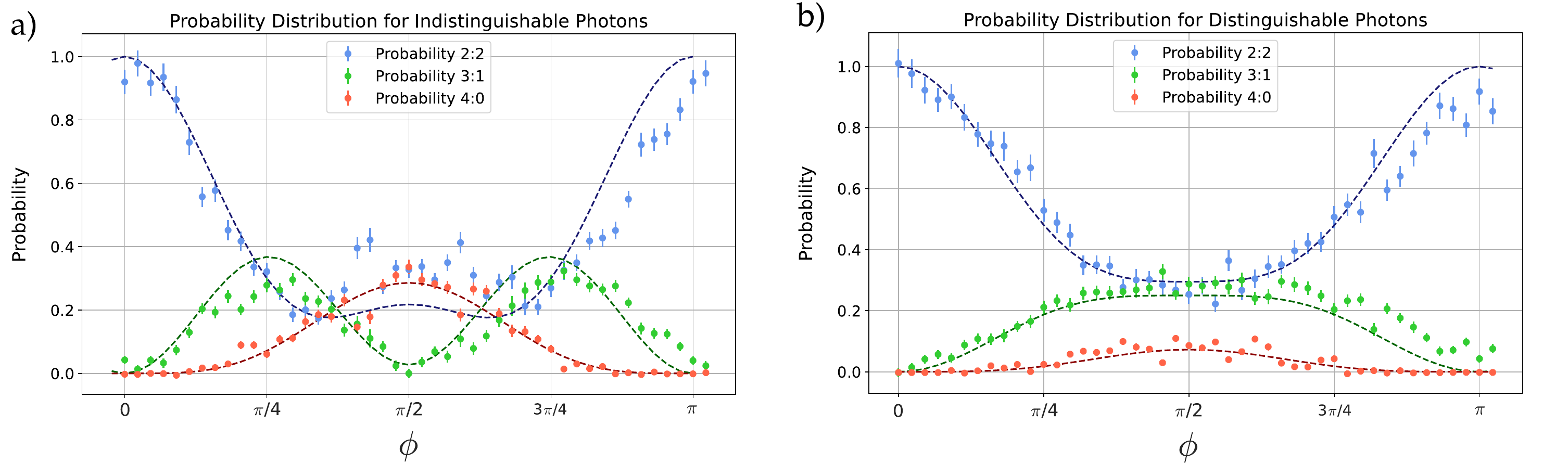}
    \caption{Probability distributions $P^{(4,2:2)}$ (blue), $P^{(4,3:1)}$ (green) and $P^{(4,4:0)}$ (red) for a) indistinguishable and b) distinguishable photons as a function of the phase difference $\phi$. Symbols represent the calculated probabilities from the measured count rates while the dashed lines correspond to the fitted probabilities according to Eq. \eqref{ec:probabilidad(4)-HOM pol}.}
    \label{fig:resultsProba}
\end{figure}

Measurements of the probability distributions $P^{(4,2:2)}$, $P^{(4,3:1)}$ and $P^{(4,4:0)}$ as a function of the HWP angle (phase difference) were taken for both a) indistinguishable ($\delta z=0$) and b) distinguishable ($\delta z> 60\mu m$) cases, shown in Figure \ref{fig:resultsProba}. 
Symbols represent the measured probabilities, calculated from the normalized count rates, while the dashed lines correspond to the fitted probabilities according to Eq. \eqref{ec:probabilidad(4)-HOM pol}.

\begin{figure}[h!]
    \centering
       \includegraphics[width=0.45\textwidth]{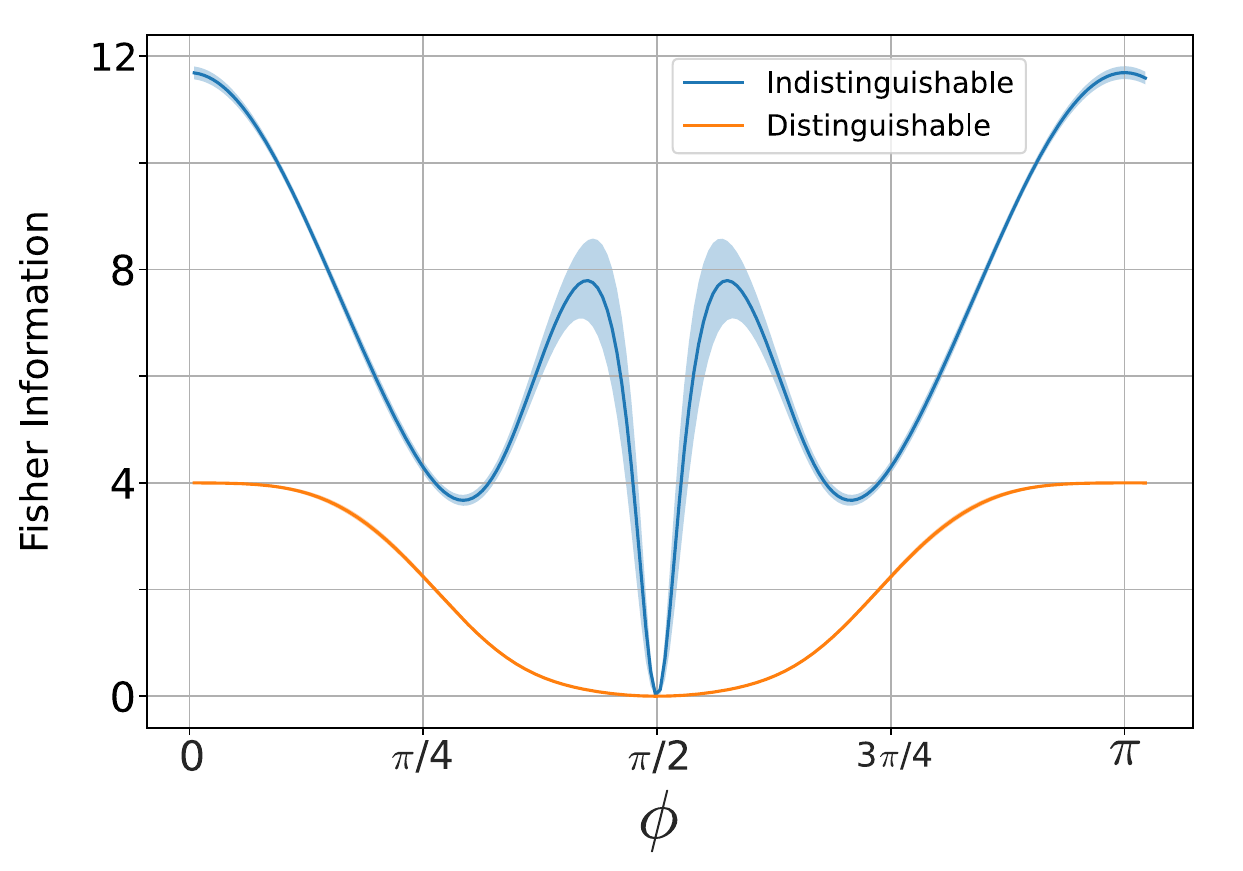}
    \caption{Fisher information for four indistinguishable (blue) and distinguishable (orange) photons as a function of the parameter $\phi$. Shaded area corresponds to the uncertainty estimated by Monte Carlo simulations (for the distinguishable case errors are smaller than the line width).}
    \label{fig:resultsFI}
\end{figure}

The parameters $\delta\lambda$ and $\delta z$ for the case of maximum indistinguishability were determined from the fitted probability distributions to be 10 nm and 10 $\mu$m, respectively. The value of $\delta\lambda$ is within the FWHM of the interference filters used in the detection system, while the value of $\delta z$ differs from the desired optimal conditions due to noise and other experimental imperfections, especially for the case of $P^{(4,2:2)}$, which shows considerable background noise that affects the visibility and sensitivity of the measurements. {Overall, the measured probability distributions are faithfully described by our theoretical model. In addition, by contemplating different degrees of freedom, our model allows for a full characterization of the experimental conditions, providing information on different parameter values.}

The Fisher information was then calculated from the fitted probability distributions as a function of the phase difference $\phi$ using Eq. \eqref{eq:FI_classic}, for both indistinguishable and distinguishable photons. The obtained values are shown in Fig. \ref{fig:resultsFI}. The shaded area corresponds to the uncertainty in the FI, calculated by Monte Carlo simulations of experimental runs with the same statistics as the measured count rates.
For both cases the FI exhibits fluctuations with respect to the parameter $\phi$. This behavior is expected given that complete indistinguishability was not achieved, as mentioned above, making the FI no longer independent of the parameter to be estimated ({compare with} Fig. \ref{fig:FI_4}). The maximum value is achieved for $\phi=0,\pi$ in both cases, obtaining $F^{4,ind}=11.6\pm0.1$ and $F^{4,dis}=3.998\pm0.002$.  
Despite fluctuations, the FI for indistinguishable photons exceeds that for distinguishable photons at all phase values, other than $\phi = \pi/2$.
{For quadrature phase angles we observe an almost three-fold increase in the achievable precision when using fully indistinguishable photons as the resource, and, most notably, an order of magnitude improvement for the precision in the determination of phase angles is expected at a wide range around the $\pi/2$ value.}

\section{Concluding remarks}
\label{sec:conc}

We presented theoretical and experimental results for the problem of parameter estimation using a four-photon Holland-Burnett probe state. We have described the correlation functions needed to calculate the probability distributions associated with our experimental scheme.
These were calculated as a function of different degrees of freedom, which contribute to the indistinguishability of the photons. Our characterization allows a complete description of the attainable precision in the problem of parameter estimation using this multi-photon state as the input probe in a HOM experiment.

For the case of completely indistinguishable photons, the
probability distributions show significant variations in the phase, presenting non-monotonic behaviors. 
{In a theoretical context, our proposed measurement scheme (defined by the input state and the POVM) achieves the maximum precision}, given by the QFI, for all phase values: $F(\phi)=F_Q(\phi)=12$.
As the path length difference increases {adding distinguishability to the system}, 
the interference patterns exhibit fewer fluctuations with the phase, suggesting that the system becomes less sensitive to changes in $\phi$. As a consequence, the FI decreases and becomes dependent with the phase. This behavior is observed experimentally, where complete indistinguishability is hard to achieve. Nevertheless, a maximum precision given by $F^{4,ind}=11.6\pm0.1$ was calculated for the experimentally generated indistinguishable four-photon state, surpassing the SQL estimated from the distinguishable photon case were we obtained $F^{4,dis}=3.998\pm0.002$.

Our results contribute to the understanding of multi-particle interference and its impact in the problem of parameter estimation.
The general framework presented in this work is of great relevance to photonic quantum metrology where {both photon interference and indistinguishability} are crucial resources to access higher precisions than those determined by classical states. 
{The flexibility in choosing parameters in our description allows for better control over experimental conditions and the possibility to adjust those that enable a significant improvement in the precision obtained in the estimation process. Our model can significantly contribute to the design of phase estimation schemes in quantum metrology with higher sensitivity by accessing different degrees of freedom of the photons. }

\appendix
\section{Probability distributions}
\label{app:A}

In order to calculate each of the probabilities in Eq. \eqref{ec:probabilidad(4)-HOM pol} the specific fourth order correlation functions need to be obtained.
These functions are, in turn, determined by the specific input states and the detection configuration, which define a particular temporal correlation operator $\hat A$.

{
Before calculating these joint probabilities, we will first introduce the photon representation as wave packet operators with a finite frequency bandwidth $\Delta$. When describing any continuous-mode state that contains a finite number of photons it is convenient to use Gaussian wave packets centered at a frequency $\omega_0$, whose spectral amplitude can be expressed as:
\begin{equation}
    \Phi(\omega)=\left(\frac{1}{2\pi\Delta^2}\right)^{1/4}\exp\left[-\frac{(\omega_0-\omega)^2}{4\Delta^2}-i\left(\omega_0-\omega\right)\delta\tau_0\right].
    \label{ec: espectro gaussiano freq}
\end{equation}

Then, by using the continuous mode creation and annihilation operators $\hat{a}^\dag(\omega)$ and $\hat{a}(\omega)$, the photon wave packet operators can be defined, where the creation operator for photons in mode $\Phi(\omega)$ is expressed as
\begin{equation}
    \hat a^\dag_{\Phi}=\int d\omega \Phi(\omega)\hat{a}^\dag(\omega).
    \label{ec: creacion espectro}
\end{equation}

In the narrow-bandwidth approximation ($\Delta << \omega_0$) we can define the wave packet amplitude as the Fourier transform of Eq. \eqref{ec: espectro gaussiano freq}: 
\begin{equation}
        \xi(t)=\frac{1}{\sqrt{2\pi}}\int d\omega\,e^{-i\omega t}\Phi(w)=
    \left(\frac{2}{\pi t_c^2}\right)^{1/4}\exp\left[-\frac{(t-\delta\tau_0)^2}{t_c^2} -i\omega_0(t-\delta\tau_0)\right]
\label{ec:gauss temp}
\end{equation}
which corresponds to a spatiotemporal Gaussian mode with $t_c=1/\Delta$ the coherence time describing the wave packet dispersion.

By defining the Fourier transformed operators of $\hat {a}^\dag(\omega)$ and $\hat {a}(\omega)$, creation and annihilation operators can also be assigned to spatiotemporal modes:
\begin{equation}
        \hat{a}^\dag(t)=\frac{1}{\sqrt{2\pi}}\int d\omega\,e^{-i\omega t}\hat {a}^\dag(\omega),\quad
        \hat{a}(t)=\frac{1}{\sqrt{2\pi}}\int d\omega\,e^{i\omega t}\hat {a}(\omega).\\
\end{equation}

In the same way as for Eq. \eqref{ec: creacion espectro} we can define the creation operator of a photon wave packet in spatiotemporal mode $\xi(t)$ as
\begin{equation}
        \hat a^\dag_{\xi}=\int dt \xi(t)\hat{a}^\dag(t).
    \label{ec: creacion tiempo}
\end{equation}
}

These operators will allow us to obtain the explicit form of the measured probabilities by integrating Eq. \eqref{ec.probabilidad cuatro fotones-HOM camino}.  As an example, in such representation, a photon state $\ket{\varphi}$ of two photons, one in mode 
 $\xi_1$ with polarization $s_1$ and the other in mode $\xi_2$ with polarization $s_2$ can be written as
\begin{equation}
\ket{\varphi}=\ket{1_{\xi_1}}_{s_1}\ket{1_{\xi_2}}_{s_2}\propto \hat{a}^\dag_{\xi_1} \hat{a}^\dag_{\xi_2}\ket{0}.
    \label{example_state}
\end{equation} 

Using this formalism, we subsequently show how to obtain the probability of detecting four photons with the same polarization $P^{(4,4:0)}$.
The corresponding four-photon temporal correlation operator, assuming $s_3$ output polarization is therefore:

\begin{equation}
\hat{A}_{s3,s3,s3,s3}^{(4:0)}(t_1,t_2,t_3,t_4)=\frac{1}{\sqrt{4!}}\hat{a}_{s3}^{\dagger}(t_1)\hat{a}_{s3}^{\dagger}(t_2)\hat{a}_{s3}^{\dagger}(t_3)\hat{a}_{s3}^{\dagger}(t_4)\frac{1}{\sqrt{4!}}\hat{a}_{s3}(t_4)\hat{a}_{s3}(t_3)\hat{a}_{s3}(t_2)\hat{a}_{s3}(t_1)
\label{A_4phout}
\end{equation}

According to Eq. \eqref{ec:HWP}, the operator for the $s_3$ output mode can be expressed in terms of the input modes as 
\begin{equation}
\hat{a}_3^{\dagger}(t_i)= -i \left[ \cos(2\theta)\hat{a}_1^{\dagger}(t_i)+\sin(2\theta)\hat{a}_2^{\dagger}(t_i) \right],
\label{outin_a3}
\end{equation}
where the ``$s$'' label has been omitted for simplicity of notation. 
Consequently we can express $\hat A$ in terms of the input operators: 
\begin{equation}
\begin{split}
\hat{A}_{s3,s3,s3,s3}^{(4:0)}(t_1,t_2,t_3,t_4)=&\frac{1}{4!}  \left[ \cos(2\theta)\hat{a}_1^{\dagger} (t_1)+\sin(2\theta)\hat{a}_2^{\dagger}(t_1) \right] \times\left[ \cos(2\theta)\hat{a}_1^{\dagger}(t_2)+\sin(2\theta)\hat{a}_2^{\dagger}(t_2) \right] \\
& \times\left[ \cos(2\theta)\hat{a}_1^{\dagger}(t_3)+\sin(2\theta)\hat{a}_2^{\dagger}(t_3) \right]\times\left[ \cos(2\theta)\hat{a}_1^{\dagger}(t_4)+\sin(2\theta)\hat{a}_2^{\dagger}(t_4) \right] \\ 
&\times\Big[ \cos(2\theta)\hat{a}_1(t_4)+\sin(2\theta)\hat{a}_2(t_4) \Big]\times\Big[\cos(2\theta)\hat{a}_1(t_3)+\sin(2\theta)\hat{a}_2(t_3)\Big] \\
& \times \Big[\cos(2\theta)\hat{a}_1(t_2)+\sin(2\theta)\hat{a}_2(t_2)\Big]\times \Big[\cos(2\theta)\hat{a}_1(t_1)+\sin(2\theta)\hat{a}_2(t_1) \Big].
\end{split}
\label{AxBxCxD}
\end{equation}

Upon distribution of all the products of sums, it must be noticed that, since the input state is a two-photon state on each mode, 
\begin{equation}
\ket{\Phi^\inp}=\ket{2_{\xi_1}}_{1}\ket{2_{\xi_2}}_{2}\propto \left(\hat{a}^\dag_{\xi_1}\right)^2 \left(\hat{a}^\dag_{\xi_2}\right)^2\ket{0},
\label{inputstates}
\end{equation}
several terms of Eq. \eqref{AxBxCxD} will vanish; namely all the terms involving three or more creation or destruction operators. For this particular selection of input and output states the temporal correlation operator takes the form
\begin{equation}
\begin{split}
\hat{A}_{s3,s3,s3,s3}^{(4:0)}(t_1,t_2,t_3,t_4)=\frac{\cos^4(2\theta)\sin^4(2\theta)}{24} &\Big[
\hat{a}_1^{\dagger}(t_1)\hat{a}_1^{\dagger}(t_2) \hat{a}_2^{\dagger}(t_3)\hat{a}_2^{\dagger}(t_4)+ 
\hat{a}_2^{\dagger}(t_1)\hat{a}_2^{\dagger}(t_2) \hat{a}_1^{\dagger}(t_3)\hat{a}_1^{\dagger}(t_4) \\
&+\hat{a}_2^{\dagger}(t_1)\hat{a}_1^{\dagger}(t_2) \hat{a}_1^{\dagger}(t_3)\hat{a}_2^{\dagger}(t_4)+
\hat{a}_2^{\dagger}(t_1)\hat{a}_1^{\dagger}(t_2) \hat{a}_2^{\dagger}(t_3)\hat{a}_1^{\dagger}(t_4)\\
& +\hat{a}_1^{\dagger}(t_1)\hat{a}_2^{\dagger}(t_2) \hat{a}_1^{\dagger}(t_3)\hat{a}_2^{\dagger}(t_4)+
\hat{a}_1^{\dagger}(t_1)\hat{a}_2^{\dagger}(t_2) \hat{a}_2^{\dagger}(t_3)\hat{a}_1^{\dagger}(t_4)   \Big] \\
& \times\Big[\hat{a}_1(t_4)\hat{a}_1(t_3) \hat{a}_2(t_2)\hat{a}_2(t_1)+ \hat{a}_1(t_4)\hat{a}_2(t_3) \hat{a}_1(t_2)\hat{a}_2(t_1) \\
&+\hat{a}_1(t_4)\hat{a}_2(t_3) \hat{a}_2(t_2)\hat{a}_1(t_1)+
\hat{a}_2(t_4)\hat{a}_1(t_3) \hat{a}_1(t_2)\hat{a}_2(t_1)\\
& +\hat{a}_2(t_4)\hat{a}_1(t_3) \hat{a}_2(t_2)\hat{a}_1(t_1)+
\hat{a}_2(t_4)\hat{a}_2(t_3) \hat{a}_1(t_2)\hat{a}_1(t_1)   \Big].
\end{split}
\label{A_4phout_2pairin}
\end{equation}

The corresponding fourth order correlation function can be calculated as
\begin{equation}
   G^{(4)}(t_1,t_2,t_3,t_4)= \bra{\Phi^\inp}\hat{A}_{s3,s3,s3,s3}^{(4:0)}(t_1,t_2,t_3,t_4) \ket{\Phi^\inp};
\end{equation}
when the product of sums in Eq. \eqref{A_4phout_2pairin} is distributed, a sum of 36 terms is obtained. Each of these terms $g$ has the form of
\begin{equation}
    g=\frac{\cos^4(2\theta)\sin^4(2\theta)}{24} \cdot \bra{2_{\xi_1}}_1\bra{2_{\xi_2}}_2 \hat{a}_i^{\dagger}(t_1)\hat{a}_j^{\dagger}(t_2) \hat{a}_k^{\dagger}(t_3)\hat{a}_l^{\dagger}(t_4)\hat{a}_{i'}(t_4)\hat{a}_{j'}(t_3) \hat{a}_{k'}(t_2)\hat{a}_{l'}(t_1)\ket{2_{\xi_1}}_{1}\ket{2_{\xi_2}}_{2}
\end{equation}
where $\{i,j,k,l\}$ and $\{i',j',k',l'\}$ run over the six permutations of the sequence $\{1,1,2,2\}$. Using standard properties of number states and creation and destruction operators and re-ordering the product of scalar terms, each term $g$ of the sum  can be written in terms of the temporal wavepackets of each input mode, and we obtain an expression that is separable in the variables $t_1$, $t_2$, $t_3$ and $t_4$.  
\begin{equation}
    g=\frac{4\cos^4(2\theta)\sin^4(2\theta)}{24}  
    \xi_i^*(t_1)\xi_{l'}(t_1)\cdot\xi_j^*(t_2)\xi_{k'}(t_2) 
    \cdot \xi_k^*(t_3)\xi_{j'}(t_3)\cdot\xi_l^*(t_4)\xi_{i'}(t_4).
    \label{single_G_term}
\end{equation}

The four-photon detection probability $P^{(4,4:0)}$ can be calculated, in the limit of the detection time $\Delta t_0$ much larger than the coherence time of the wavepacket $t_c$, by integrating the fourth order correlation function $G^{(4)}$: 
\begin{equation}
    P^{(4)}=\iiiint dt_1dt_2dt_3dt_4\,G^{(4)}(t_1,t_2,t_3,t_4).
    \label{ec.probabilidad cuatro fotones-HOM camino app}
\end{equation}
It should be noted that permutation of the mode indexes in Eq. \eqref{single_G_term} leads to different expressions depending on whether the modes on the products $\xi_i^*(t)\xi_j(t)$ are similar or not. The expression for $P^{(4,4:0)}$ can then be rearranged as
\begin{equation}
\begin{split}
   P^{(4,4:0)}=& \frac{\cos^4(2\theta)\sin^4(2\theta)}{6}\left\{ 6 \, \left (\int \left | \xi_1(t)\right |^2 dt \right)^2 \cdot \left (\int \left | \xi_2(t)\right |^2 dt \right)^2 \right.\\
   & +6\,  \Re \left (\int \left | \xi_1(t)\right |^2 dt \cdot\int \left | \xi_2(t)\right |^2 dt \cdot \int \xi^*_1(t)\xi_2(t) dt \cdot \int \xi^*_2(t)\xi_1(t) dt \right) \\
   &\left. +24\,  \Re \left [ \left (\int \xi^*_1(t)\xi_2(t) dt \right)^2 \cdot \left (\int \xi^*_2(t)\xi_1(t) dt\right)^2 \right]\right\}.
\end{split}
   \end{equation}

This expression allows for the calculation of the probability of obtaining a four-photon output at one of the outputs as a function of several parameters of the photon wavepackets, such as the temporal width, the central frequency and the arrival time of the input states. We assume Gaussian temporal wavepackets for both input modes of the form
\begin{equation}
    \xi_k(t)=\left (\frac{2}{\pi t_c^2}\right)^{1/4}\exp{\left[-\frac{\left(t-\delta\tau_k\right)^2}{t_c^2}-i\omega_k\left(t-\delta\tau_k \right)\right]} 
\end{equation}
 where $t_c$ is the coherence time of the wavepackets, which is inversely proportional to their bandwidth, $\delta\tau_1=-\delta\tau_2\equiv\delta\tau$ are the arrival times of (the two) input modes, $\omega_k$ is the central optical frequency of each mode and the frequency difference is $\delta\omega=(\omega_2-\omega_1)/2$. The probability of obtaining four photons at a single output is therefore
 \begin{equation}
    P^{(4,4:0)}(2\theta, t_c,\delta\tau,\delta \omega)=\cos^4(2\theta)\sin^4(2\theta)\left(1+ e^{-\frac{8\delta\tau^2}{ t_c^2}-2\,\delta \omega^2 t_c^2}+4e^{-\frac{4\delta\tau^2}{ t_c^2}-\delta \omega^2 t_c^2}\right).
\end{equation}

In order to calculate the probabilities of the other detection configurations at the output, namely $P^{(4,3:1)}$ and $P^{(4,2:2)}$, the correlation operator has to be modified accordingly: 
\begin{equation}
\hat{A}_{s3,s3,s3,s4}^{(3:1)}(t_1,t_2,t_3,t_4)=\frac{1}{\sqrt{3!}}\hat{a}_{3}^{\dagger}(t_1)\hat{a}_{3}^{\dagger}(t_2)\hat{a}_{3}^{\dagger}(t_3)\hat{a}_{4}^{\dagger}(t_4)\frac{1}{\sqrt{3!}}\hat{a}_{4}(t_4)\hat{a}_{3}(t_3)\hat{a}_{3}(t_2)\hat{a}_{3}(t_1) 
\label{A_31phout}
\end{equation}
and
\begin{equation}
\hat{A}_{s3,s3,s4,s4}^{(2:2)}(t_1,t_2,t_3,t_4)=\frac{1}{\sqrt{2!}\sqrt{2!}}\hat{a}_{3}^{\dagger}(t_1)\hat{a}_{3}^{\dagger}(t_2)\hat{a}_{4}^{\dagger}(t_3)\hat{a}_{4}^{\dagger}(t_4)\frac{1}{\sqrt{2!}\sqrt{2!}}\hat{a}_{4}(t_4)\hat{a}_{4}(t_3)\hat{a}_{3}(t_2)\hat{a}_{3}(t_1) 
\label{A_31phout}
\end{equation}
respectively; the operator for the $s_4$ output mode is 
\begin{equation}
\hat{a}_4^{\dagger}(t_i)= -i \left[ \sin(2\theta)\hat{a}_1^{\dagger}(t_i)-\cos(2\theta)\hat{a}_2^{\dagger}(t_i) \right].
\label{outin_a4}
\end{equation}


\end{document}